\documentclass[twocolumn,tighten,times]{aastex631}

\usepackage[T1]{fontenc}

\newcommand\ergs{erg~s$^{-1}$}
\newcommand\ergcms{erg~cm$^{-2}$~s$^{-1}$}
\newcommand\kms{km~s$^{-1}$}

\newcommand{\hbeta}{H{$\beta$}}
\newcommand{\halpha}{H{$\alpha$}}
\newcommand{\OIII}{[O\,{\sc iii}]}
\newcommand{\OIIIa}{[O\,{\sc iii}]\,$\lambda$4959}
\newcommand{\OIIIb}{[O\,{\sc iii}]\,$\lambda$5007}
\newcommand{\OIIIwave}{[O\,{\sc iii}]\,$\lambda\lambda$4959,5007}

\newcommand{\OI}{[O\,{\sc i}]}
\newcommand{\OIwave}{[O\,{\sc i}]\,$\lambda$6300}
\newcommand{\NII}{[N\,{\sc ii}]}
\newcommand{\NIIwave}{[N\,{\sc ii}]\,$\lambda$6583}
\newcommand{\SII}{[S\,{\sc ii}]}
\newcommand{\SIIwave}{[S\,{\sc ii}]\,$\lambda\lambda$6716,\,6731}
\newcommand{\SIIa}{[S\,{\sc ii}]\,$\lambda$6716}
\newcommand{\SIIb}{[S\,{\sc ii}]\,$\lambda$6731}

\newcommand{\HII}{H\,{\sc ii}}
\newcommand{\OIIItohb}{[O\,{\sc iii}]\,$\lambda$5007$/$H{$\beta$}}

\shorttitle{VLT MUSE observations of the NGC 1313 X-2 nebula}
\shortauthors{Zhou et al.}

\graphicspath{{./}{fig/}}

\begin{document}

\title{VLT MUSE observations of the bubble nebula around NGC 1313 X-2 and evidence for additional photoionization}

\author{Changxing Zhou}
\affiliation{Department of Engineering Physics, Tsinghua University, Beijing 100084, China}

\author{Fuyan Bian}
\affiliation{European Southern Observatory, Alonso de C\'{o}rdova 3107, Casilla 19001, Vitacura, Santiago 19, Chile}

\correspondingauthor{Hua Feng}
\email{hfeng@tsinghua.edu.cn}

\author[0000-0001-7584-6236]{Hua Feng}
\affiliation{Department of Astronomy, Tsinghua University, Beijing 100084, China}
\affiliation{Department of Engineering Physics, Tsinghua University, Beijing 100084, China}

\author{Jiahui Huang}
\affiliation{Department of Engineering Physics, Tsinghua University, Beijing 100084, China}

\begin{abstract}
The bubble nebula surrounding NGC 1313 X-2 is believed to be powered by high velocity winds from the central ultraluminous X-ray source (ULX) as a result of supercritical accretion. With the Multi-Unit Spectroscopic Explorer (MUSE) observation of the nebula, we find enhanced \OIII\ emission at locations spatially coincident with clusters of stars and the central X-ray source, suggesting that photoionization in addition to shock-ionization plays an important role in powering the nebula.  The X-ray luminosity of the ULX and the number of massive stars in the nebula region can account for the required ionizing luminosity derived with \emph{MAPPINGS V}, which also confirms that pure shocks cannot explain the observed emission line ratios. 
\end{abstract}


\section{Introduction}

Ultraluminous X-ray sources (ULXs) are extremely accreting compact objects, and occupy the high end of the luminosity function of high mass X-ray binaries \citep{Mineo2012}. They display X-ray luminosities exceeding the Eddington limit of a typical stellar mass black hole, with spectral and timing behaviors distinct from Galactic X-ray binaries \citep[for a review see][]{Kaaret2017}.  Identification of neutron stars in ULXs \citep{Bachetti2014,Fuerst2016,Israel2017,Israel2017a,Carpano2018,Sathyaprakash2019,RodriguezCastillo2020} and their ordinary spectral properties \citep{Pintore2017,Walton2018} suggest that supercritical accretion occurs in the majority of these systems.  

Due to the presence of strong radiation pressure, winds are expected to launch under supercritical accretion \citep{Shakura1973,Meier1982,Lipunova1999,King2003,Poutanen2007,Shen2016}. This is also predicted by numerical simulations \citep{Ohsuga2011,Jiang2014,Hashizume2015,Sadowski2016,Takahashi2016,Abarca2018,Kitaki2018,Kitaki2021}, which all reveal that there is a central funnel where ultra-fast winds (0.1--0.4~$c$) are launched.  Observationally, the wind is evidenced with the detection of high velocity absorption features in the X-ray spectra of ULXs \citep{Pinto2016,Pinto2017,Pinto2021,Walton2016,Kosec2018}. 

Interaction of the wind with the interstellar medium (ISM) may produce shock-ionized bubble nebulae, which have been found around some ULXs \citep{Pakull2002,Pakull2003,Ramsey2006,Abolmasov2007,Abolmasov2008,Russell2011,Soria2021}.  These bubbles have a size of some 100~pc, expanding at a velocity of about 100~km~s$^{-1}$.  Based on the emission line luminosity, bubble size, and expansion velocity, the mechanical power of the wind or jet that drives the bubble is estimated to be about $10^{39}$--$10^{40}$~\ergs, suggesting that the accretion is supercritical \citep{Pakull2003,Abolmasov2007}, in agreement with the detection of diffuse X-ray emission in one of them \citep{Belfiore2020}.

NGC 1313 X-2 is a prototype ULX among the first discovered \citep{Colbert1995}.  A bubble nebula is found to be coincident with the X-ray source position and has been extensively studied \citep{Pakull2002, Pakull2003,Zampieri2004,Mucciarelli2005,Pakull2006,Ramsey2006,Ripamonti2011}.  The nebula is elongated with a geometry of 590~pc $\times$ 410~pc ($26\arcsec \times 18\arcsec$).  The presence of strong \SII\ and \OI\ emission lines and the supersonic expansion suggests that the nebula is shock-ionized \citep{Pakull2002,Pakull2003}, while the most likely powering source is the wind/jet from the central ULX \citep{Pakull2008} with a mechanical power estimated to be about $1.5 \times 10^{39}$~\ergs~\citep{Pakull2002}. 

In this paper, we present a spatially resolved spectroscopic study of the bubble nebula around NGC 1313 X-2 using the MUSE instrument \citep{Bacon2010} on the Very Large Telescope (VLT).  The observations are described in \S~\ref{sec:obs}, the data analysis and results are presented in \S~\ref{sec:res}, and the physical nature is discussed in \S~\ref{sec:dis}. We assume a distance of 4.6~Mpc to NGC 1313 based on Cepheids measurements~\citep{Qing2015}.

\section{Observations}
\label{sec:obs} 

The VLT/MUSE observations of NGC1313 X-2 were performed on the nights of 2019 October 16 and November 12 (UT). The seeing during the observations was from $0\farcs8$ to $1\farcs0$, and sky transparency conditions were clear on 2019 October 16 and partially cloudy on 2019 November 12. The MUSE WFM-NOAO-E mode was used to cover the wavelength between 4600\AA\ and 9350\AA. The observations include four 1400-second exposures: two of them with a position angle of 0\arcdeg\ and the rest two with an angle of 90\arcdeg. 

The data were reduced by the EsoRex pipeline version 3.13.2 \citep{Weilbacher2020}. We used the {\it muse\_scibasic} recipe to correct bias frames, lamp flats, arc lamps, twilight flats, geometry, and illumination exposures for individual exposures. Then the {\it muse\_scipost} recipe was used for telluric and flux calibrations with the standard star taken in the same night. At last the individual exposures were aligned using the {\it muse\_exp\_align} recipe and combined to generate the final datacube using the {\it muse\_exp\_combine} recipe. The field of view of the final datacube is $1\arcmin \times 1\arcmin$ and the spaxel scale is $0\farcs2 \times 0\farcs2$. 

\section{DATA ANALYSIS AND RESULTS}
\label{sec:res}

\begin{figure}
\centering
\includegraphics[width=\columnwidth]{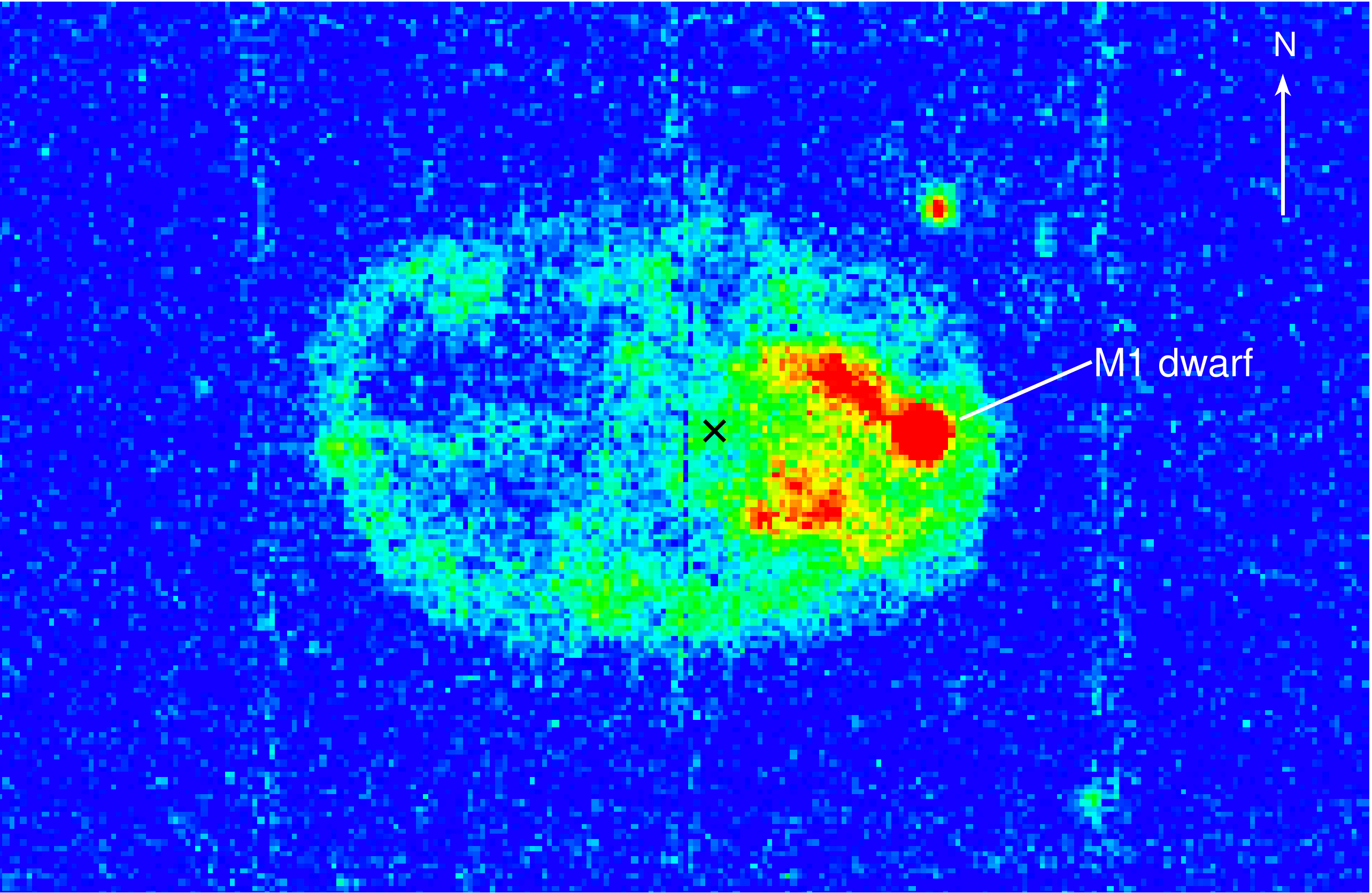}
\includegraphics[width=\columnwidth]{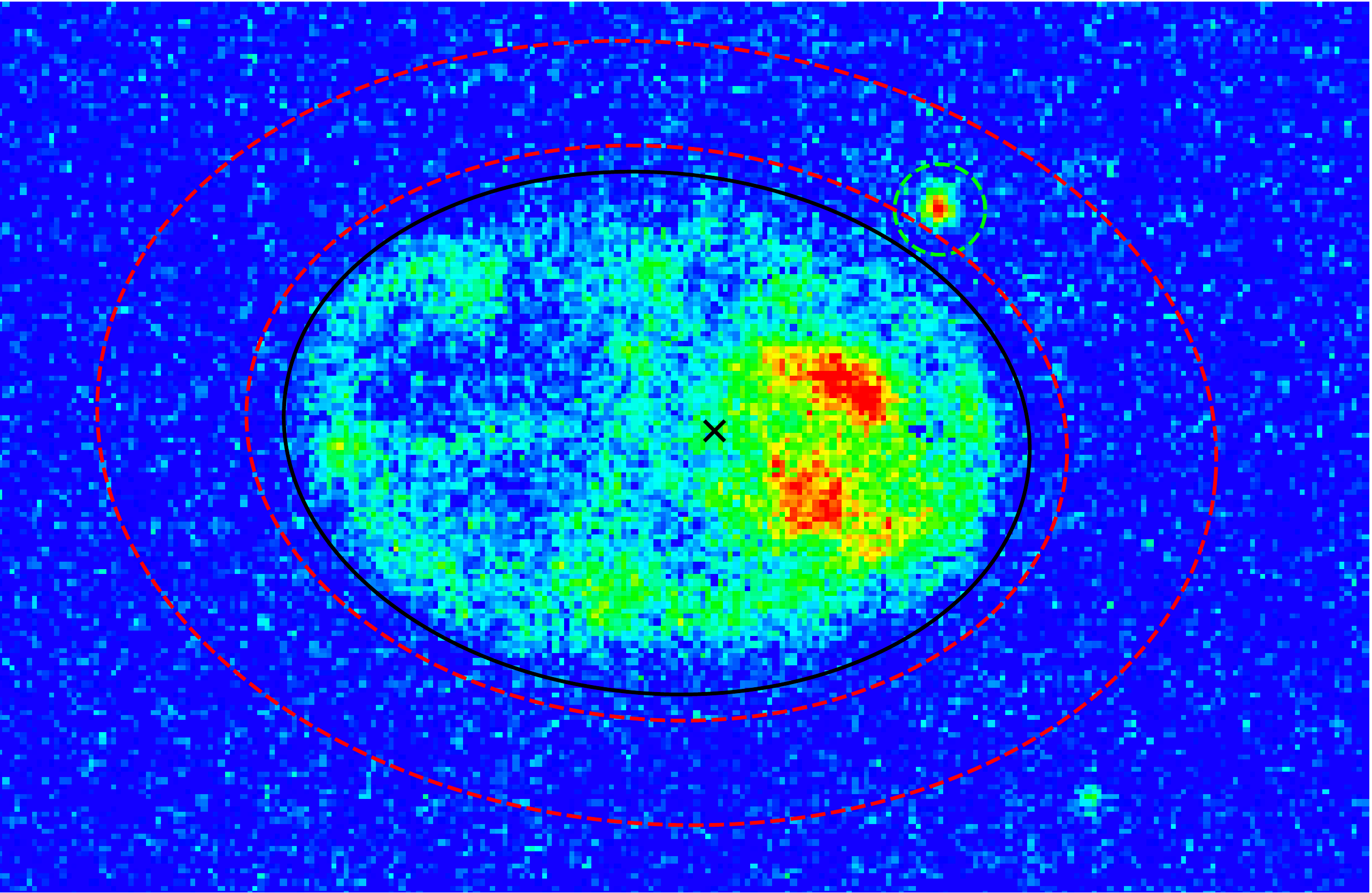}
\caption{MUSE \halpha\ ($\pm 5$\AA\ around the line centroid) images around NGC 1313 X-2 with ({\bf top}) and without ({\bf bottom}) the continuum emission. The cross marks the X-ray position of NGC 1313 X-2~\citep{Liu2007}. A foreground M1 dwarf appears along the line of sight of the nebula and is visible only with the continuum emission. The black ellipse defines the nebula region and the red dashed lines excluding a green dashed circle encircle the background region. The arrow points north and has a length of 5\arcsec\ (112~pc).}
\label{fig:halpha}
\end{figure}

Figure~\ref{fig:halpha} shows the nebula images in the \halpha\ band, one with line plus continuum emission and the other with line emission only.  A bright object appears near the west end of the nebula, and is identified to be a foreground M1 dwarf at a distance of 750~pc \citep{GaiaCollaboration2021} along the line of sight of the nebula \citep[also see][]{Ramsey2006}.  

\begin{figure*}
\centering
\includegraphics[width=\textwidth]{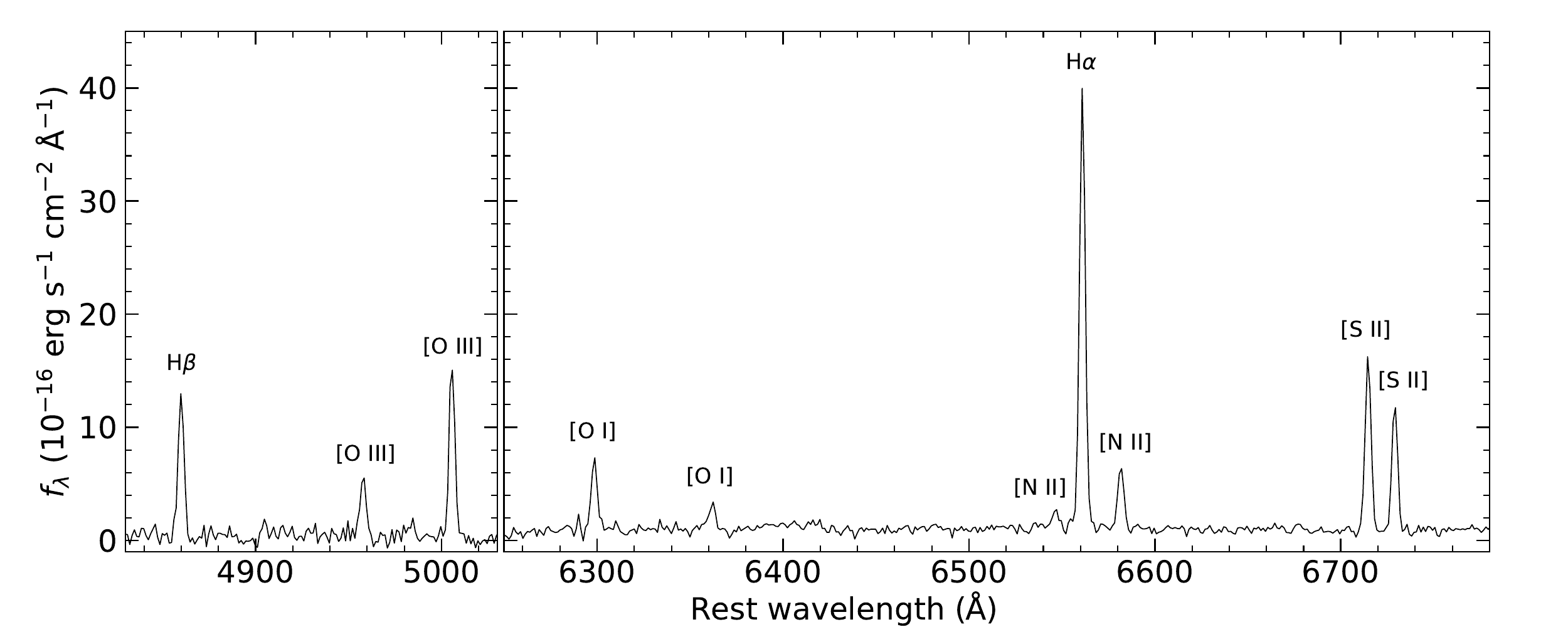}
\caption{The MUSE spectra in the blue (left) and red (right) bands of the NGC 1313 X-2 nebula. }
\label{fig:spec}
\end{figure*}

We find that the nebula spectra are contaminated by unresolved stellar emission in the host galaxy projected on the nebula region \citep[see the Hubble images in][]{Grise2008}.  The stellar emission contributes a continuum component with absorption lines, while the nebula spectrum mainly consists of emission lines.  In order to remove the stellar contamination, we extract a template spectrum from a background region defined in Figure~\ref{fig:halpha}.  For the nebula, we are only interested in the emission line properties in two bands, a blue band that contains \hbeta\ and \OIII, and a red band with \OI, \halpha, \NII, and \SII. We extracted the nebula spectrum from an elliptical region defined in Figure~\ref{fig:halpha}, and fit it with the template in the two bands, respectively, at wavelengths excluding emission lines. In the fitting, we employ a scale factor and a shift on the flux onto the template spectrum.  Then the best-fit stellar template is removed from the nebula spectrum. The blue and red bands are not fitted jointly because different scales are needed.  We have to assume that the stellar contamination is insensitive to the sky position, because the statistics from a small number of pixels does not allow us to perform the template fitting.   The spectrum extracted from a single pixel after subtracting the stellar template exhibits fluctuations consistent with the statistical error, justifying the validity of the technique.  The final spectra in the blue and red bands are shown in Figure~\ref{fig:spec}. 

Each emission line is fitted locally with a Gaussian to a continuum subtracted spectrum. The continuum is determined with a linear function from nearby wavelengths free of the emission component. The Balmer lines are fitted independently. Either the \OIIIwave\ or \SIIwave\ doublet are fitted jointly, imposing the same velocity and width; for \OIIIwave, we further constrain the flux ratio to be 2.98 \citep{Storey2000}, which is in good agreement with the ratio derived from independent measurements.  The line flux, velocity relative to the host galaxy \citep[$z = 0.001568$ or $v = 470$~\kms;][]{Koribalski2004}, and the full-width-half-maximum (FWHM) velocity dispersion corrected for instrument line broadening\footnote{Available in Section 3.2 of the MUSE user manual, see \url{https://www.eso.org/sci/facilities/paranal/instruments/muse/doc/ESO-261650\_MUSE\_User\_Manual.pdf}. The lines seen in nearby \HII\ regions are found to have widths consistent with the instrument line broadenings.} are derived for bright lines and listed in Table~\ref{tab:line} for the whole nebula.  The line fluxes derived from the observation on October 16 are systematically higher than those on November 12 because of different sky conditions, and have been corrected as in the clear sky condition. 

The average radial velocity of the whole nebula is around $-80$~\kms\ (negative means a blue shift) with respect to the bulk velocity of 470~\kms\ as mentioned above. They are consistent with the line velocities measured from three nearby, possibly \HII\ regions. For example, one of them marked by a dashed green circle in Figure~\ref{fig:halpha} has a velocity of  $-81 \pm 2$~\kms\ measured with \halpha\ or $-80 \pm 2$~\kms\ with \SII. Thus, we adopt $-80$~\kms\ as the systematic velocity. This confirms the conclusion in \citet{Pakull2002} that the bubble nebula is indeed associated with the host galaxy. 

\begin{deluxetable}{lcllc}
\tabletypesize\scriptsize
\tablecaption{The emission line flux, velocity, and velocity dispersion for the whole nebula.}
\label{tab:line}
\tablewidth{\columnwidth}
\tablecolumns{5}
\tablehead{
\colhead{line} & \colhead{$f$} & \colhead{$v$} & \colhead{FWHM} & \colhead{FWHM$_{\rm inst}$} \\
\colhead{} & \colhead{($10^{-16}$~\ergcms)}& \colhead{(\kms)} & \colhead{(\kms)} & \colhead{(\kms)} 
}
\startdata
\hbeta  & $62.0 \pm 2.9$  & $-87.7 \pm 3.2$ &  $126 \pm 13$ & $177 \pm 6$ \\
\OIIIa    & $26.4 \pm 1.1$  & $-72.8 \pm 2.9$ &  $129 \pm 11$  & $173 \pm 6$ \\
\OIIIb    & $78.7 \pm 3.3$   & $-72.8 \pm 2.9$ & $129 \pm 11$ & $171 \pm 6$ \\
\OIwave & $32.8 \pm 1.0$   & $-87.4 \pm 1.6$ & $121 \pm 6$ & $126 \pm 4$ \\ 
{[O\,{\sc i}]\,$\lambda$6364} & $11.4 \pm 2.2$   & $-88 \pm 11$ & $128 \pm 37$ & $124 \pm 4$ \\ 
{[N\,{\sc ii}]\,$\lambda$6548} & $8.9 \pm 1.5$   & $-72 \pm 8$ & $117 \pm 33$ & $119 \pm 4$ \\ 
\halpha & $191.5 \pm 2.1$  & $-83.7 \pm 0.6$ &  $110 \pm 2$ & $119 \pm 4$ \\
\NIIwave & $29.4 \pm 0.7$  & $-80.8 \pm 1.3$ &  $127 \pm 4$ & $119 \pm 4$ \\ 
\SIIa & $76.6 \pm 1.6$  & $-80.3 \pm 1.0$ &  $110 \pm 3$ & $116 \pm 4$ \\ 
\SIIb & $55.0 \pm 1.3$   & $-80.3 \pm 1.0$ & $110 \pm 3$ & $115 \pm 4$ \\ 
\enddata
\tablecomments{$v$ is the relative velocity with respect to the host galaxy. FWHM has been corrected for instrument line broadening (FWHM$_{\rm inst}$). }
\tabletypesize\small
\end{deluxetable}

\begin{figure}[t]
\centering
\includegraphics[width=\columnwidth]{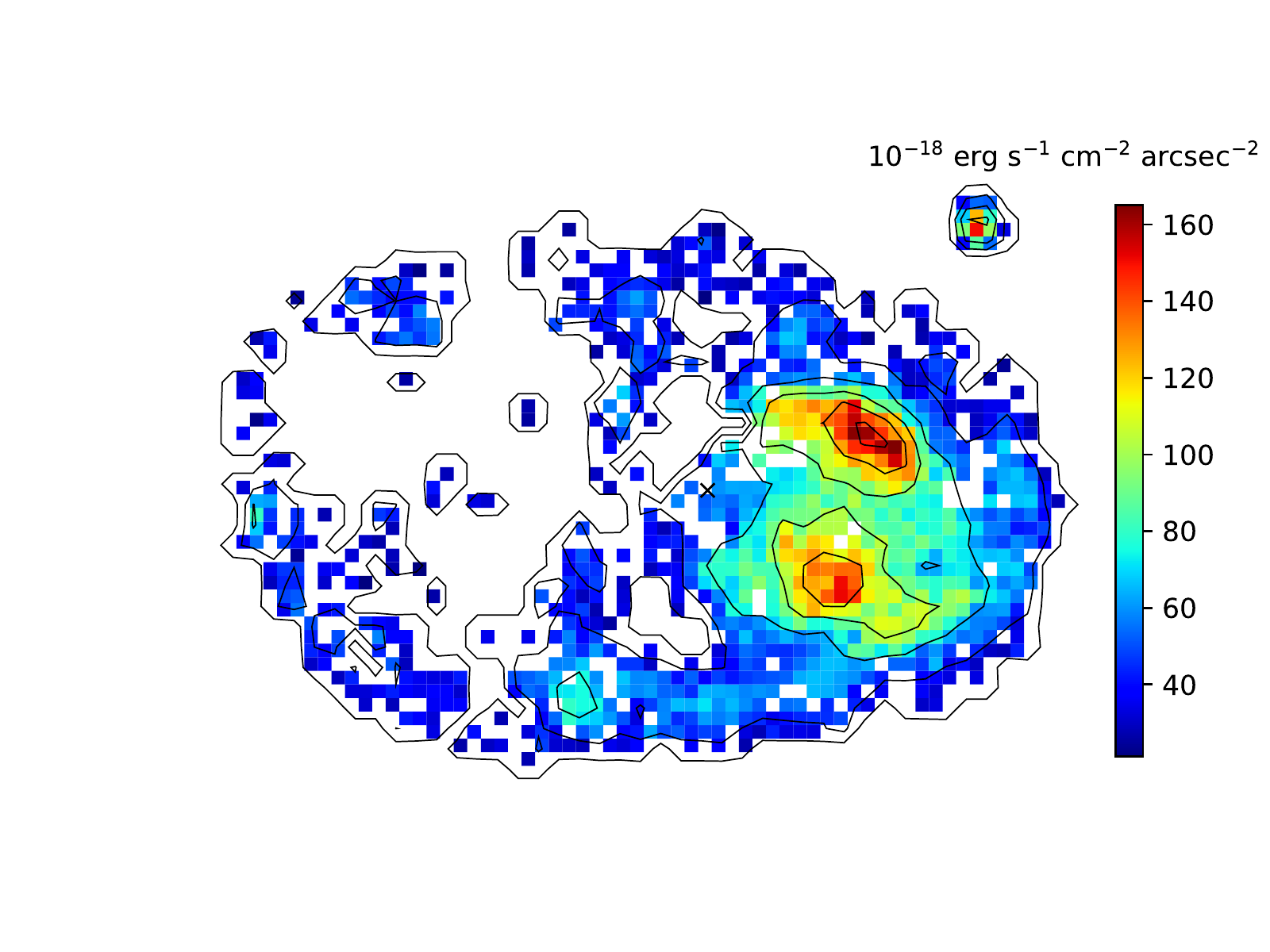}
\includegraphics[width=\columnwidth]{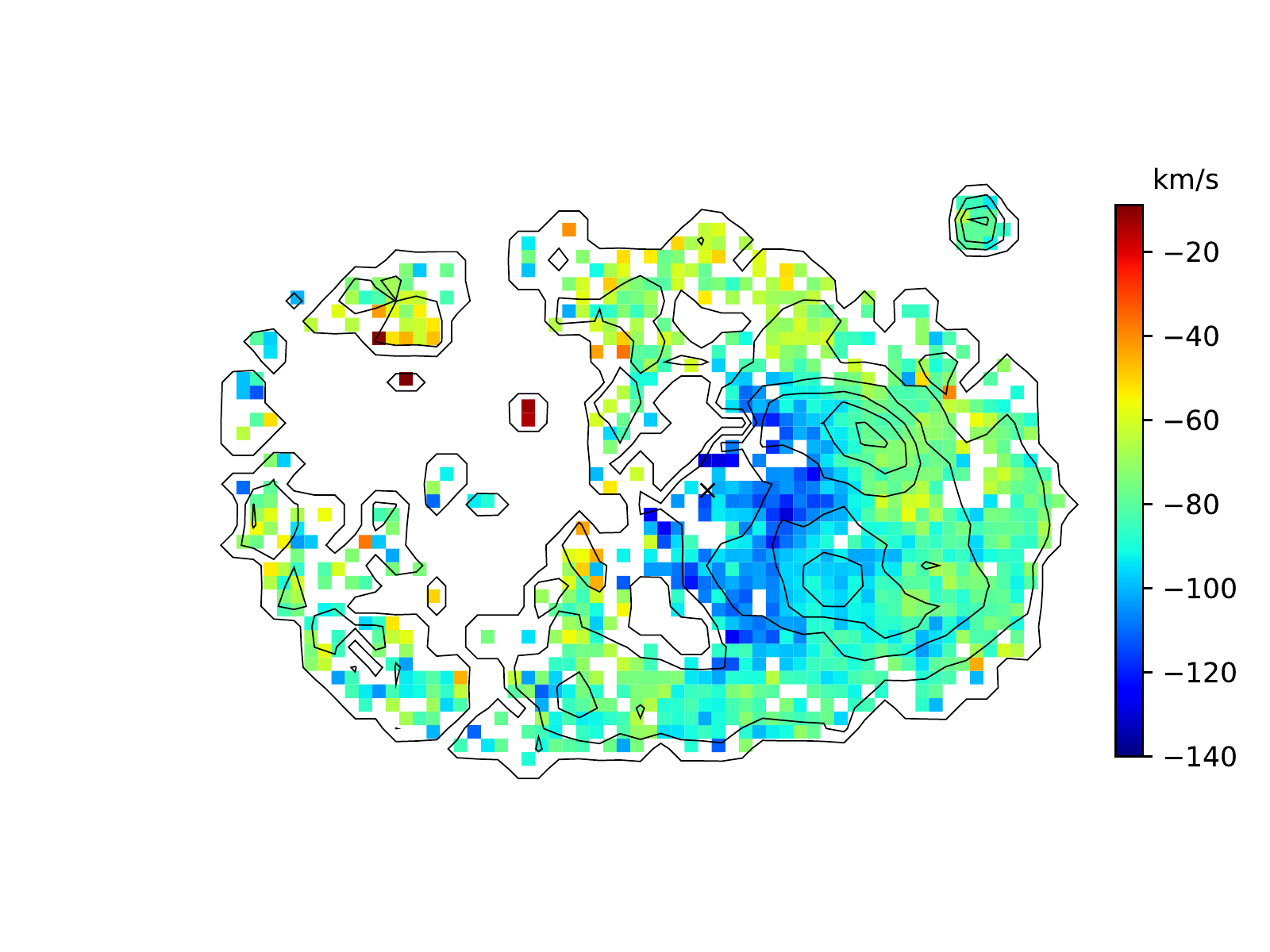}
\includegraphics[width=\columnwidth]{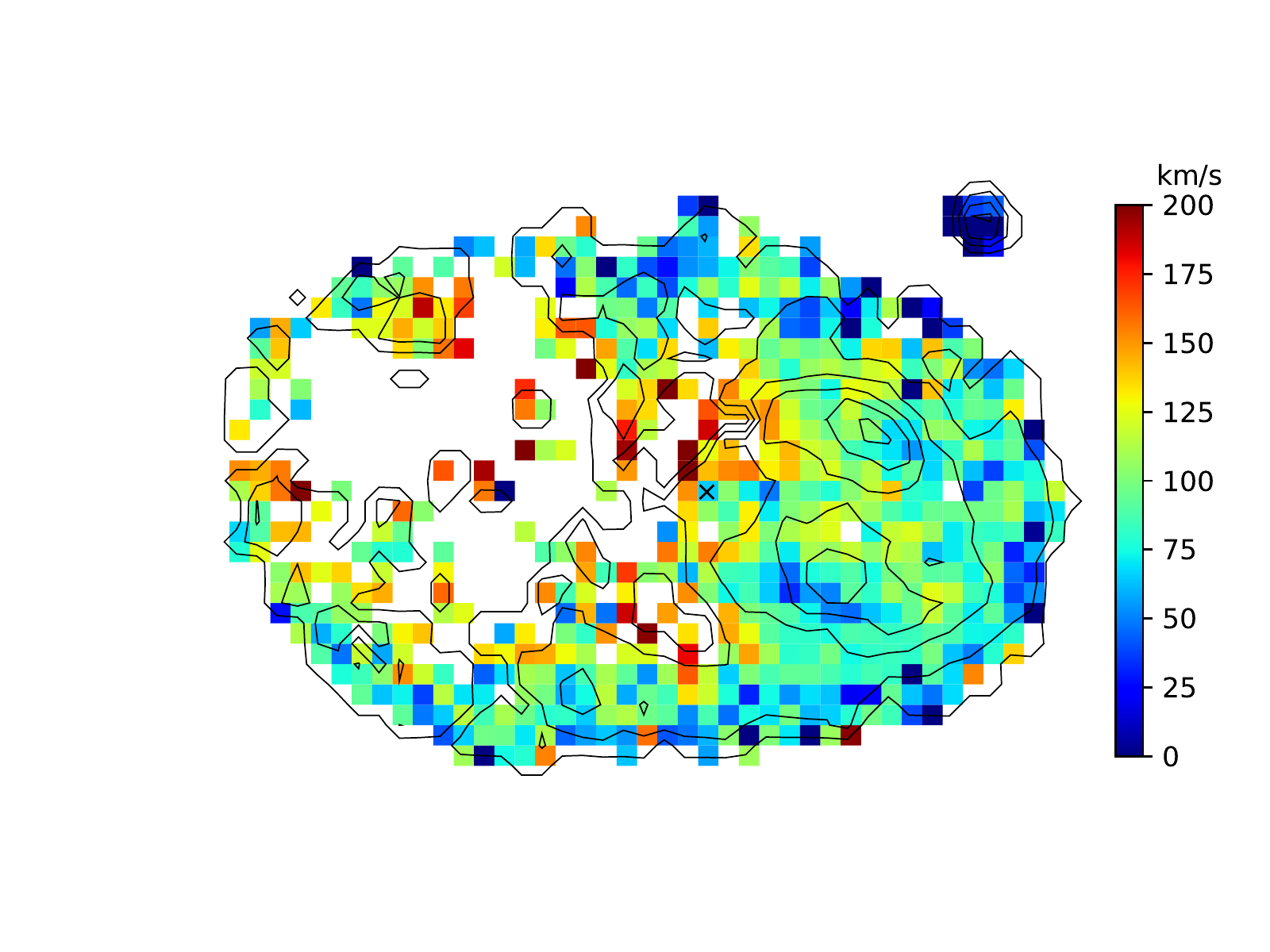}
\caption{\halpha\ flux ({\bf top}), radial velocity ({\bf middle}), and FWHM velocity dispersion ({\bf bottom}) maps of the NGC 1313 X-2 nebula, on top of the flux contours. The former two have a binning of $2 \times 2$ original pixels and the last one is binned by $3 \times 3$ pixels before spectral fitting.  Pixels are not shown if the emission line cannot be determined at a significance of 2$\sigma$. The position of the X-ray source is indicated by a cross.}
\label{fig:map}
\end{figure}

To investigate the spatial distribution of the line properties, we produce the flux, radial velocity, and FWHM velocity dispersion (instrument broadening removed) maps in Figure~\ref{fig:map}. \halpha\ is the only emission line that has a signal-to-noise ratio high enough for this purpose. We employ a $2 \times 2$ pixel binning for the flux and velocity maps, and a $3 \times 3$ binning for the FWHM map to improve the statistics.  

\begin{figure}[t]
\centering
\includegraphics[width=\columnwidth]{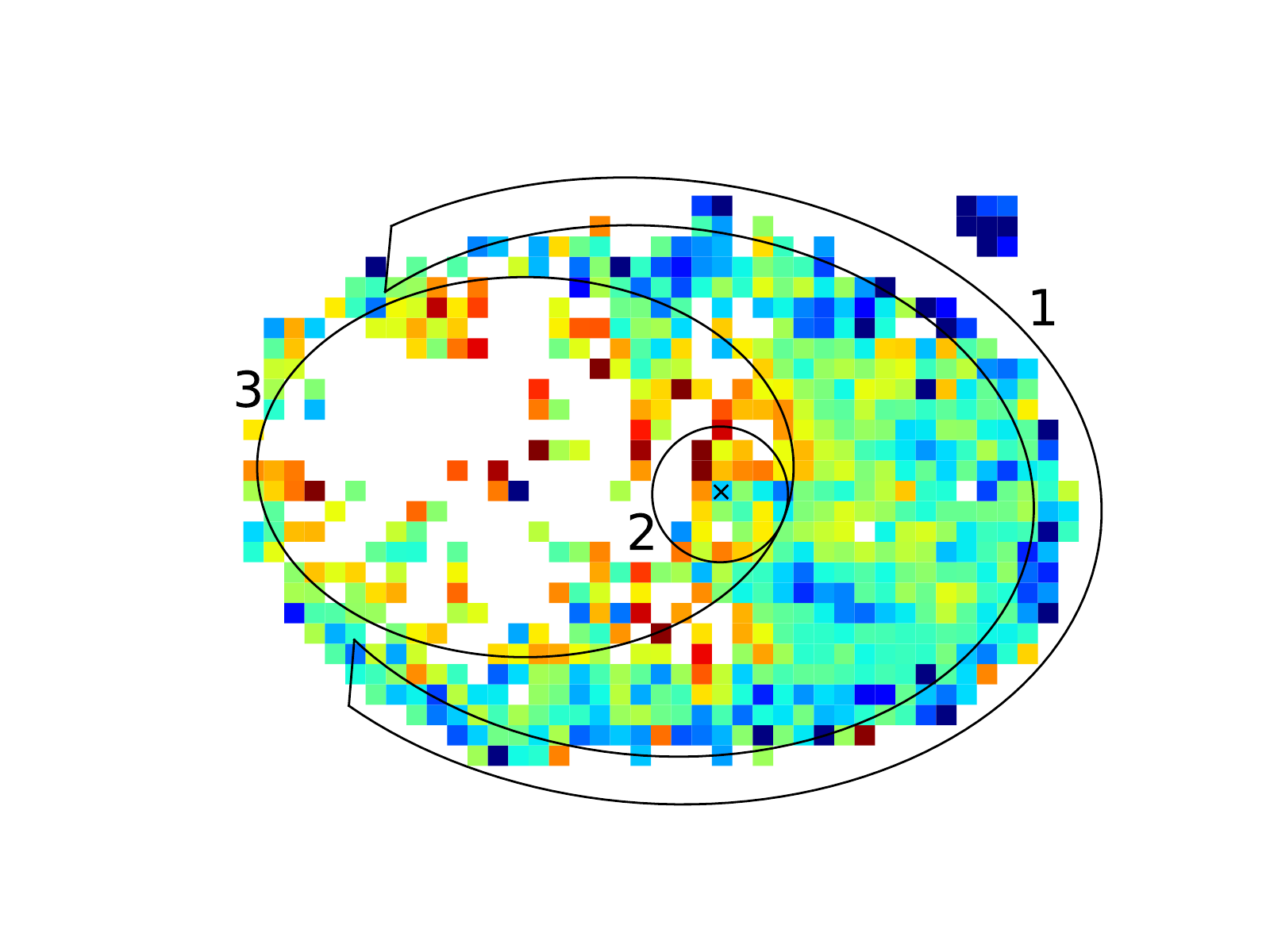}
\caption{Three regions on the \halpha\ FWHM map.  Region 1 is a 2\arcsec\ wide elliptical annulus referring to the edge of the bubble nebula excluding the part overlapped with region 3. Region 2 is a 2\arcsec-radius circle centered on the ULX position. Region 3 (with region 2 exclucded) indicates a flux cavity. }
\label{fig:3reg}
\end{figure}

We mark three particular regions on the map (see Figure~\ref{fig:3reg}).  Region 1 is a 2\arcsec\ wide elliptical annulus along the edge of the bubble, excluding the eastern end where it overlaps with region 3. This is the outermost region of the bubble, and represents the part with the lowest FWHM. Region 2, i.e., the ULX region, is a 2\arcsec-radius circle around the X-ray position, which is blue shifted compared with other regions.  Region 3 is a low flux cavity in the eastern part of the bubble. The velocity and and velocity dispersion measured with \halpha, \SII, and \OIII\ in the three regions are listed in Table~\ref{tab:3reg}. 

\begin{deluxetable}{lrr}
\tablecaption{Velocities and velocity dispersions in the three regions defined in Figure~\ref{fig:3reg} for \halpha, \SII, and \OIII.}
\label{tab:3reg}
\tablewidth{\columnwidth}
\tablecolumns{3}
\tablehead{
\colhead{Region} & \colhead{$v$} & \colhead{FWHM} \\
\colhead{} & \colhead{(\kms)} & \colhead{(\kms)} 
}
\startdata
\multicolumn{3}{c}{\halpha}\\
\noalign{\smallskip}\hline\noalign{\smallskip}  
1: edge & $-81.0 \pm 0.4$ & $73 \pm 2$ \\
2: ULX &  $-106.5 \pm 1.1$ & $136 \pm 4$ \\
3: cavity & $-76.4 \pm 0.9$ & $160 \pm 3$ \\
\noalign{\smallskip}\hline\noalign{\smallskip}  
\multicolumn{3}{c}{\SIIwave}\\
\noalign{\smallskip}\hline\noalign{\smallskip}  
1: edge & $-76.6  \pm 2.1$ & $68  \pm 10$ \\
2: ULX & $-105.9 \pm 4.4$ & $149 \pm 13$ \\
3: cavity & $-74.7  \pm 3.0$ & $158 \pm 9$ \\
\noalign{\smallskip}\hline\noalign{\smallskip}  
\multicolumn{3}{c}{\OIIIwave}\\
\noalign{\smallskip}\hline\noalign{\smallskip}  
1: edge & $-81.8  \pm 6.5$ & $76  \pm 38$ \\
2: ULX & $-107.4 \pm 2.8$ & $91  \pm 14$ \\
3: cavity & $-57.2  \pm 4.1$ & $173 \pm 14$ \\
\enddata
\end{deluxetable}

\subsection{Line decomposition}

We suppose that the line kinematics seen on each pixel of the nebula is a result of both approaching and receding motions, with respect to the systematic velocity \citep[also see][]{Soria2021}. Region 1, which is along the edge of the bubble, shows a radial velocity well consistent with the systematic velocity,  and a FWHM among the lowest in the bubble, indicating that the gases along the edge may be moving transversally.  If this is the case, the FWHM in region 1 reflects the intrinsic line width of the shock, which is around $\sim$70~\kms\ (see Table~\ref{tab:3reg}). 

Due to the presence of strong low ionization forbidden lines, such as \OI, \SII\ and \NII, the shock is believed to be radiative instead of adiabatic \citep{Heng2010,Soria2021}. For radiative shocks, it is not straightforward to infer the local shock velocity from the gas velocity dispersion \citep[see Appendix A in][]{Soria2021}. Following their recipe, which is based on the assumption of a uniformly expanding thin shell, we infer a shock velocity $v_{\rm s} \approx 0.47 {\rm FWHM_2} = 64$~\kms, or $v_{\rm s} \approx 0.735 {\rm FWHM_{all}} = 81$~\kms, where FWHM$_2$ and FWHM$_{\rm all}$ are the \halpha\ FWHM in region 2 (Table~\ref{tab:3reg}) and of the whole nebula (Table~\ref{tab:line}), respectively.  Previous studies using long slit spectroscopy argued that the shock velocity of the nebula could be as high as 80~\kms\ \citep{Pakull2002} or 100~\kms\ \citep{Ramsey2006}.  Thus, in this study, we assume a shock velocity of 80~\kms.

\begin{figure}[t]
\centering
\includegraphics[width=0.8\columnwidth]{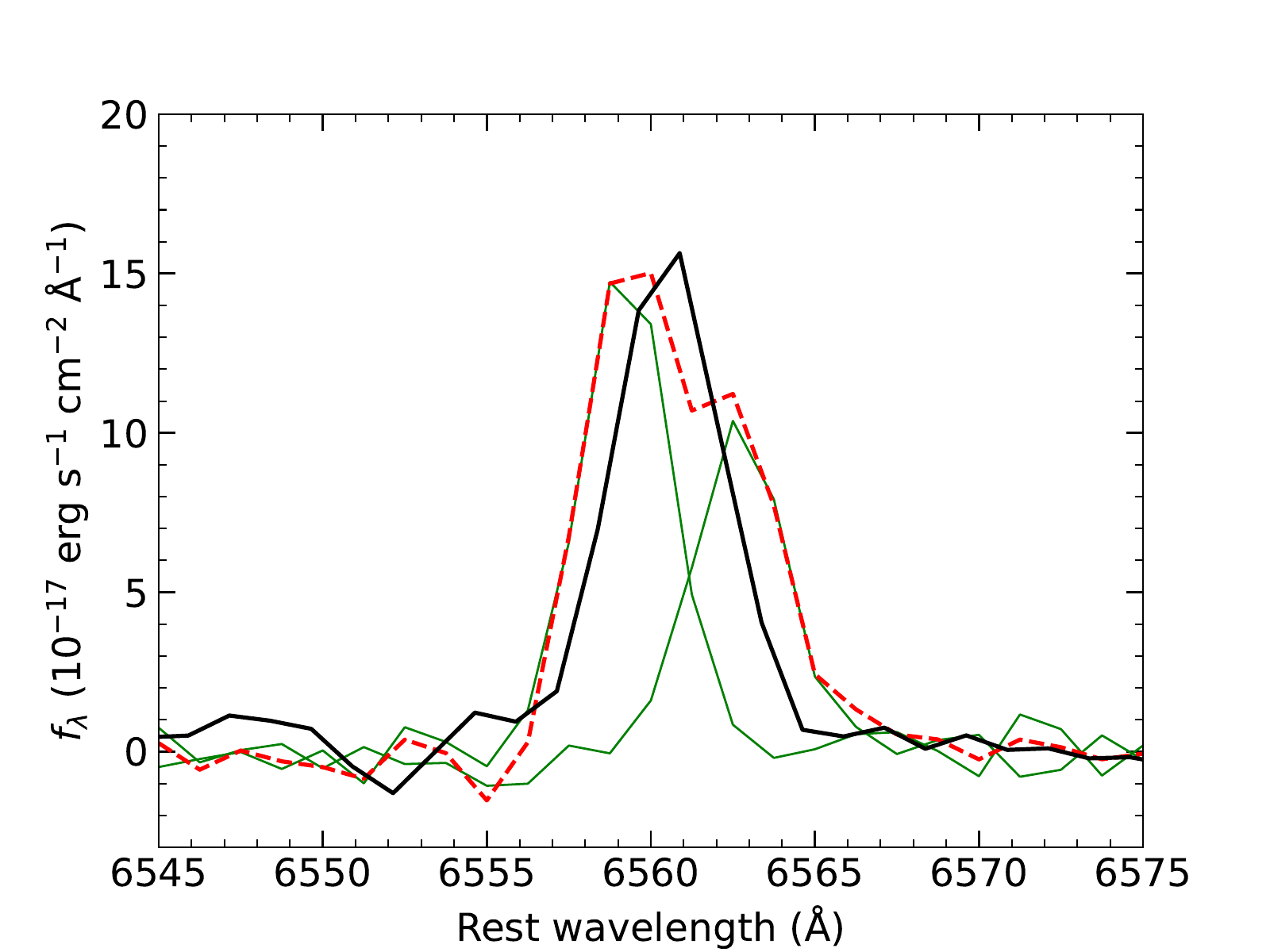}
\includegraphics[width=0.8\columnwidth]{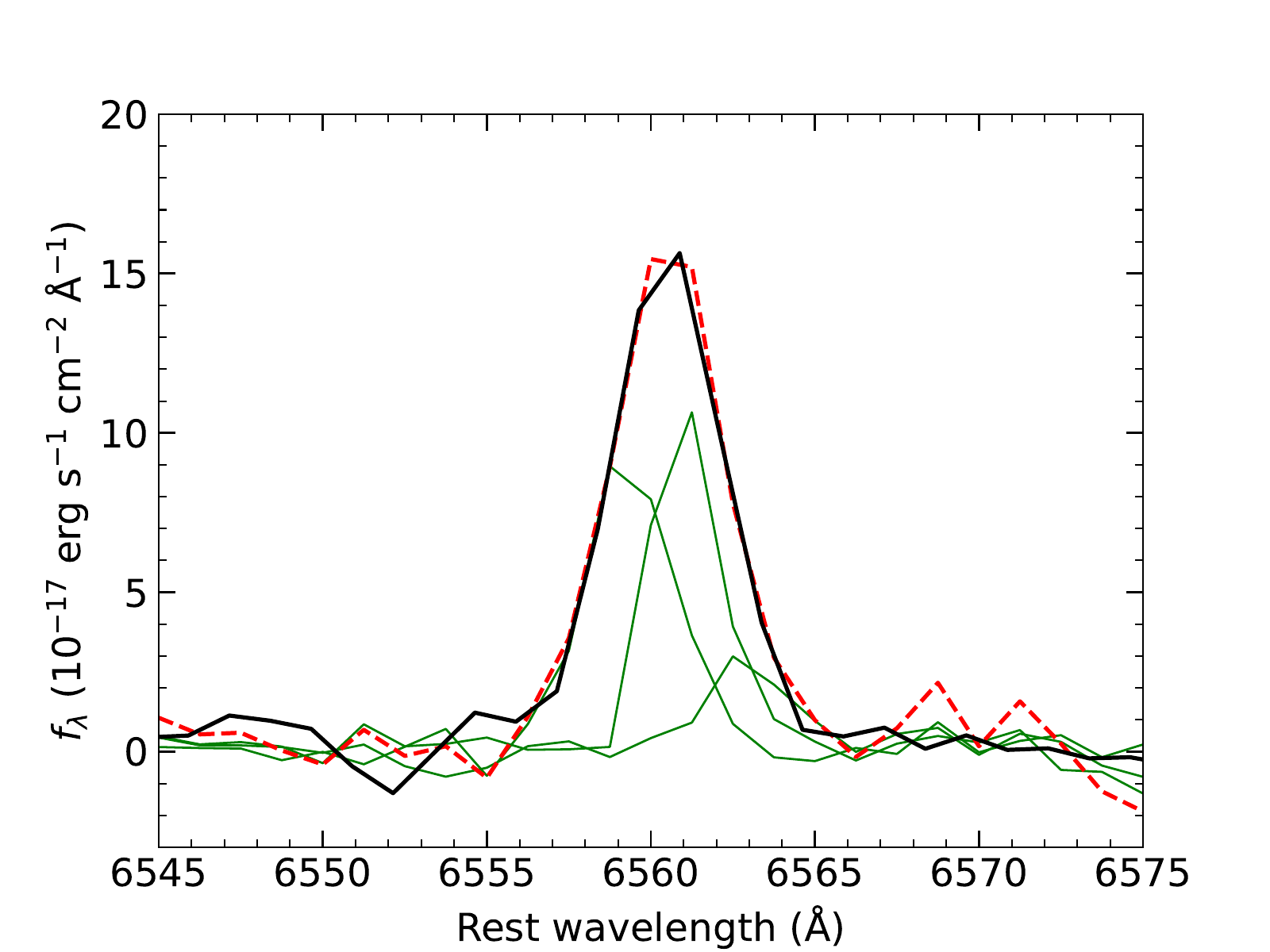}
\caption{Observed (solid black) and model (dashed red) \halpha\ emission line profiles in region 2. {\bf Top}: the model consists of two components (green), both with a FWHM of 70~\kms, approaching and receding from the systematic at a velocity of 80~\kms, respectively. {\bf Bottom}: a third, narrow component at zero-velocity is added}.
\label{fig:decomp} 
\end{figure}

Then, we try to model the observed \halpha\ line profile in the ULX region (region 2), where the bulk motion of shocked gas is radial.  Here we assume that the shock velocity equals the bulk velocity of shocked gas, which is the observed velocity, in the case of fully radiative shocks \citep{Dopita2003}.  We fix the shock velocities to be $\pm$80~\kms\ for the approaching and receding components, respectively, and assume that both have a FWHM of 70~\kms. The instrument line broadening and measurement noise are taken into account.  However, the model line width is significantly larger than the observed line width (Figure~\ref{fig:decomp}).  We obtain the same conclusions if we fix the FWHM at any value in the range of 50--100~\kms.   

This may suggest that the emission line has a more complicated velocity distribution \citep[also see the 2D spectrum in][]{Ramsey2006}. Therefore, we add a zero-velocity component with an intrinsic FWHM = 0.  The three-component model can fit the line profile adequately, with a flux ratio of 3:3:1 (Figure~\ref{fig:decomp}). The 2D spectrum in \citet{Ramsey2006} also shows a hint of a stationary component. The presence of a low-velocity narrow component could be interpreted as due to photoionization. 

\subsection{Observational evidence for additional photoionization}

As mentioned above, the strong \SII\ and \OI\ emission lines in the nebula are suggestive of predominant shock-ionization \citep{Pakull2002,Pakull2003}.  To produce high ionization emission lines like \OIII, one requires either high velocity shocks or photoionization.

\begin{figure}[t]
\centering
\includegraphics[width=0.8\columnwidth]{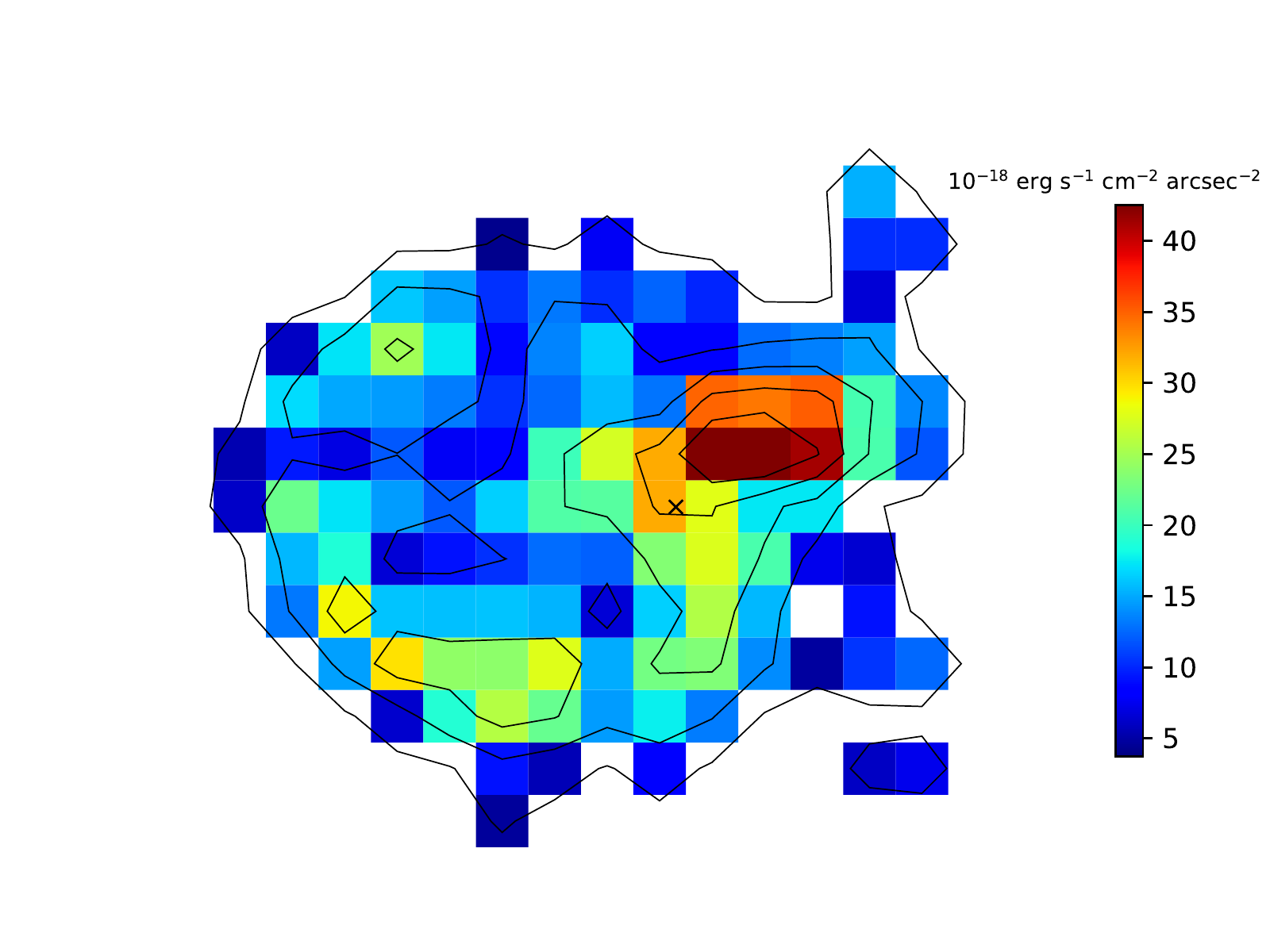}
\includegraphics[width=0.8\columnwidth]{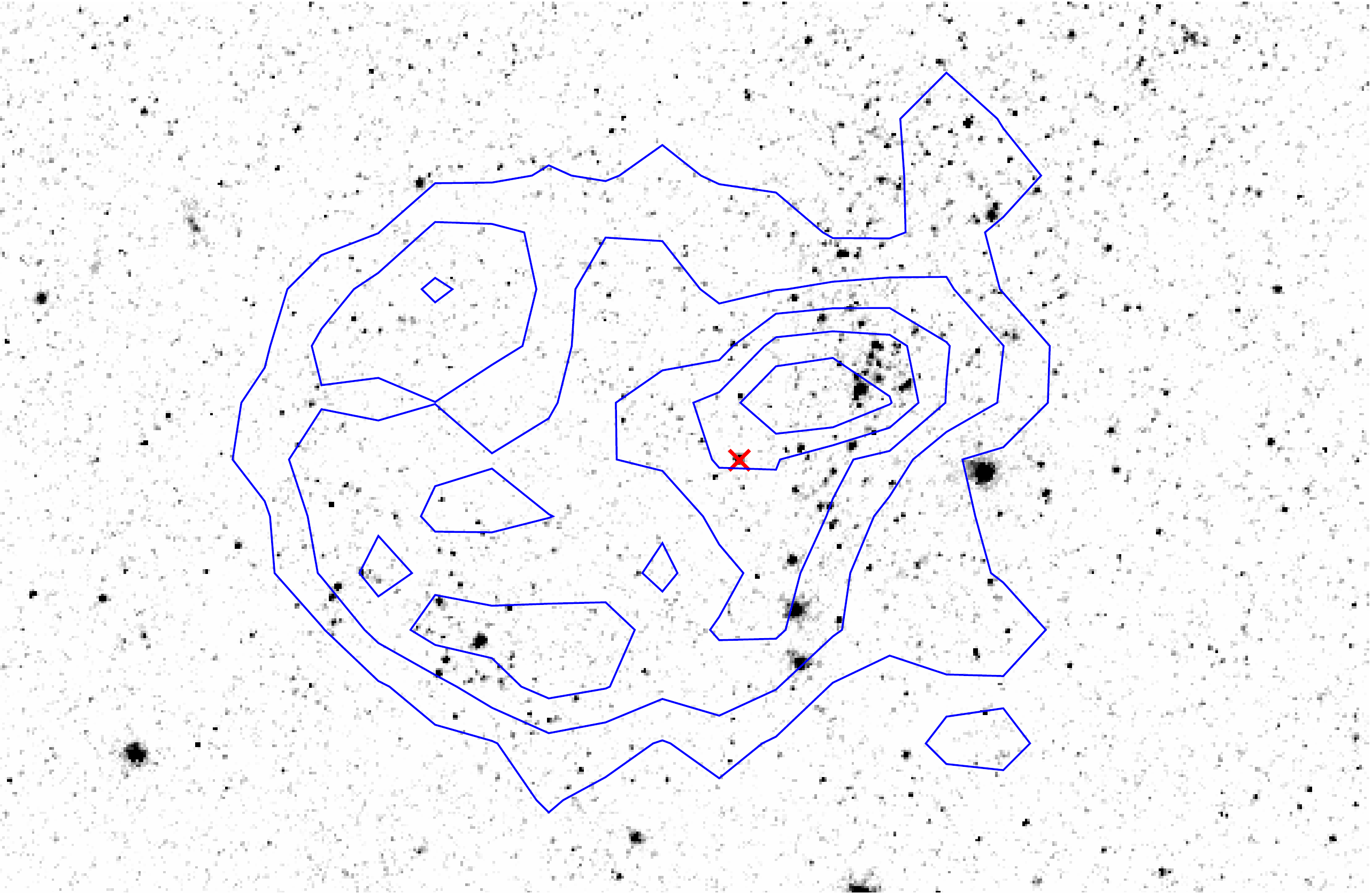}
\includegraphics[width=0.8\columnwidth]{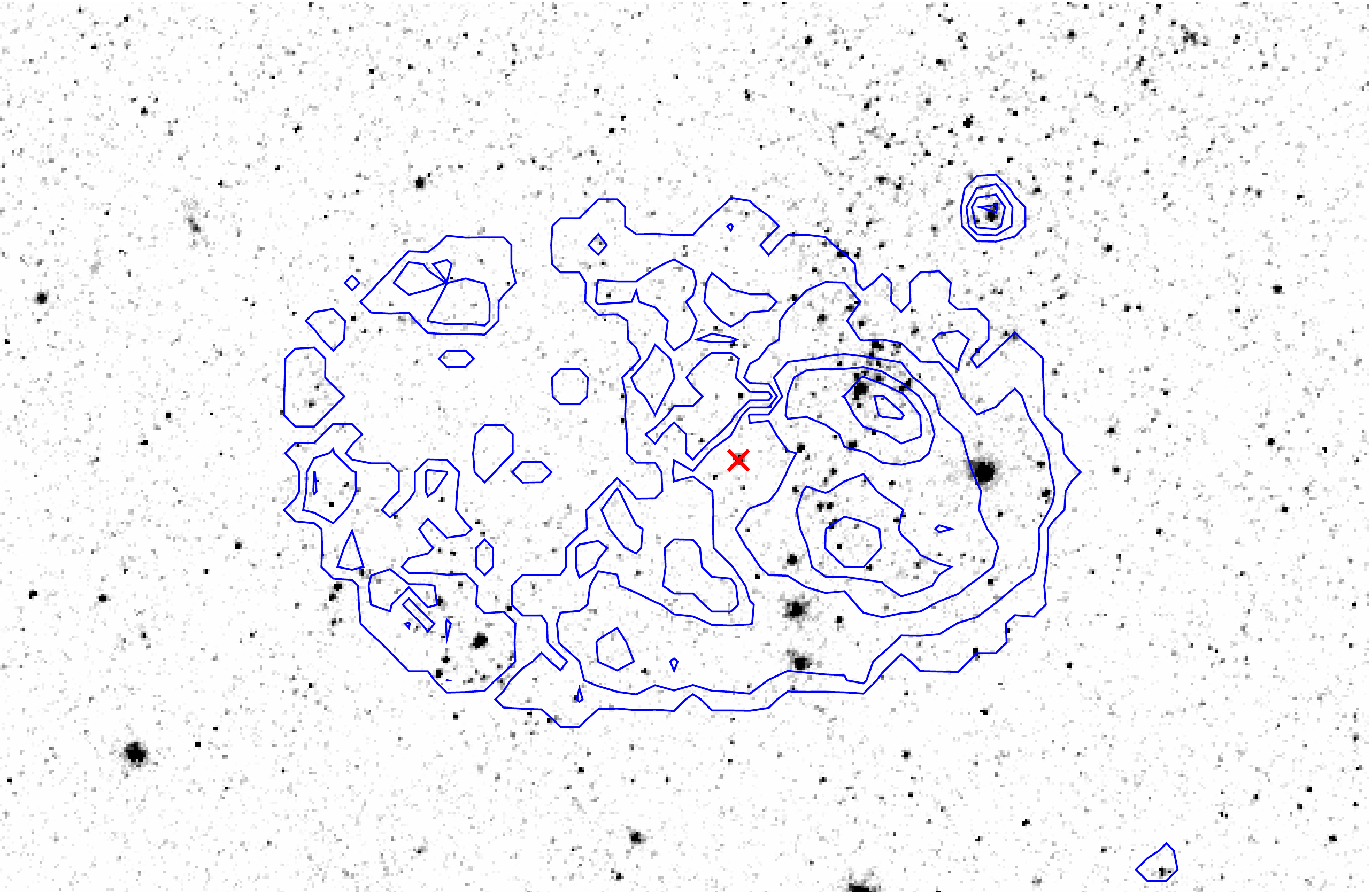}
\caption{The {\bf top} panel shows the continuum removed \OIIIb\ ($\pm 5$\AA\ around the line centroid) flux images around NGC 1313 X-2, with a binning of $9 \times 9$ pixels.  Those having a signal-to-noise ratio less than 2 are not shown. The {\bf middle} and {\bf bottom} panels display an HST ACS F435W image with \OIIIb\ or \halpha\ flux contours, respectively.  
\label{fig:oiii_hst}}
\end{figure}

We display the \OIIIb\ flux map on top of an Hubble Space Telescope (HST) ACS F435W image in Figure~\ref{fig:oiii_hst}. As one can see, regions with high \OIII\ fluxes are spatially coincident with dense stellar populations shown on the HST image.  No similar coincidence is seen with the \halpha\ map. This implies that some of the \OIII\ emission may be attributed to photoionization by massive stars.  The \OIII\ flux in the ULX region (region 2) is also enhanced, possibly caused by additional X-ray irradiation. As one can see in Table~\ref{tab:3reg}, the FWHM of \OIII\ in region 2 is significantly smaller than that of \halpha\ and \SII, implying that the \OIII\ emission is perhaps due to a different ionizing mechanism.  

\subsection{Comparison with MAPPINGS V}

To check if photoionization is needed in addition to shock-ionization, we perform simulations with \emph{MAPPINGS V} \citep{Sutherland2017}.  We assume an ISM density of 1~cm$^{-3}$, a magnetic field of 3~$\mu$G \citep{Han2017}; these two parameters are not sensitive to the results.  We also assume $Z = 0.5 Z_\sun$ as suggested by previous studies \citep{Ripamonti2011,Pintore2012} and will discuss its influence.  The models include dust calculations and allow grain destruction; these two options also have a small impact on our conclusions.  The pre-ionization state is calculated in a self-consistent manner, taking into account internal photoionization due to post-shock radiation and external photoionization from X-ray sources and O stars. 

First, we compare observations with simulations of pure shocks  (Model A), and set the shock velocity $v_{\rm s} = 80$~\kms.  However, the simulated \OIIIb\ to \hbeta\ flux ratio (on the order of $10^{-6}$) is significantly lower than that observed (see Table~\ref{tab:ratio}).  If we set a solar abundance, \OIIItohb\ is found to be 0.016, still about two orders of magnitude lower than that observed.  To match the observed \OIIItohb, one requires a shock velocity over 100~\kms\ at $0.5 Z_\sun$, or $\approx85$~\kms\ at solar abundance. In these cases, however, the \OIwave\ to \halpha\ flux ratio is found to be significantly lower than that observed, because most of the oxygens stay at a higher ionization state.  To conclude, pure shock-ionization cannot account for both \OIIItohb\ and \OIwave$/$\halpha. 

We thereby add photoionization in addition to shock-ionization (Model B). We add an X-ray source and a certain number of O stars for photoionization illuminating at a distance of 300~pc (length of the semi-major axis of the nebula). The X-ray power-law spectral index is fixed at $-1.5$, the typical value found from X-ray fitting \citep{Qiu2021}. Each O star has an effective temperature of 40000~K. We vary the X-ray luminosity and O star quantity to fit the observation, and find that an X-ray luminosity of $1.6 \times 10^{40}$~\ergs\ plus 60 O stars with a total luminosity of $6.4 \times 10^{40}$~\ergs\ can reasonably match the observations (Table~\ref{tab:ratio}).  The simulated \NIIwave$/$\halpha\ is higher than that observed; this is consistent with previous studies by \citet{Ripamonti2011} who propose that the nitrogen abundance in the nebula could be further lower. 

This ULX also displays a soft blackbody component, which could be the ionizing source. We replace the power-law component with a blackbody component and set the temperature at two extremes from observations, 0.1~keV and 0.25~keV \citep{Qiu2021}. However, it cannot fit the observed flux ratios in the luminosity range of $(0.5 - 50) \times 10^{39}$~\ergs.  If we further increase the blackbody luminosity to match the observed \OIIItohb, the predicted \SIIa$/$\halpha\ is way below that observed. 

We also perform a test with a flat ionizing spectrum ($F_\nu \propto \nu^0$) in the EUV to soft X-ray band (40~eV to 1 keV) to mimic emission from a multicolor disk with an innermost temperature of about 0.1~keV.  In this case, an X-ray luminosity of $5 \times 10^{40}$~\ergs\ plus 250~O stars can provide a good fit with the observations except \OIwave$/$\halpha, which is over-predicted by a factor of 1.6. Thus, we refer to this case as a marginal fit. 

The observed \halpha\ to \hbeta\ flux ratio is $2.87 \pm 0.14$ after correction with the Galactic extinction $E(B-V) = 0.075$ along the line of sight~\citep{Schlafly2011}. This sets the upper bound on the intrinsic \halpha$/$\hbeta.  As one can see, the pure shock model (Model A) predicts a higher \halpha$/$\hbeta, inconsistent with the observation, while the model with additional photoionization (Model B) produces  an acceptable ratio, if there is no extra reddening in the host galaxy.  This is consistent with previous studies \citep{Grise2008} that there is no extragalactic extinction for the NGC 1313 X-2 nebula. 

\begin{deluxetable}{llcc}
\tablecaption{Observed and simulated line flux ratios.}
\label{tab:ratio}
\tablewidth{\columnwidth}
\tablecolumns{4}
\tablehead{
\colhead{Line flux ratio} & \colhead{Observed} & \colhead{Model A} & \colhead{Model B} }
\startdata
 \OIIItohb & $1.269 \pm 0.079$   & 0.004    & 1.092 \\
 \OIwave$/$\halpha & $0.171 \pm 0.005$   & 0.018    & 0.201 \\ 
 \halpha$/$\hbeta & $2.868 \pm 0.138$\tablenotemark{$\ast$}   & 3.395    & 2.955 \\ 
 \NIIwave$/$\halpha & $0.154 \pm 0.004$  & 0.071    & 0.378 \\ 
 \SIIa$/$\halpha& $0.400 \pm 0.009 $  & 0.106    & 0.418 \\ 
\enddata
\tablenotetext{\ast}{The value has been corrected with the Galactic extinction.}
\tablecomments{Model A: pure shock at 80~\kms. Model B: shock at 80~\kms\ with a power-law ($F_\nu \propto \nu^{-1.5}$) X-ray source of  $1.6 \times 10^{40}$~\ergs\ and 60 O stars at 300 pc. }
\end{deluxetable}

\begin{figure}[t]
\centering
\includegraphics[width=0.8\columnwidth]{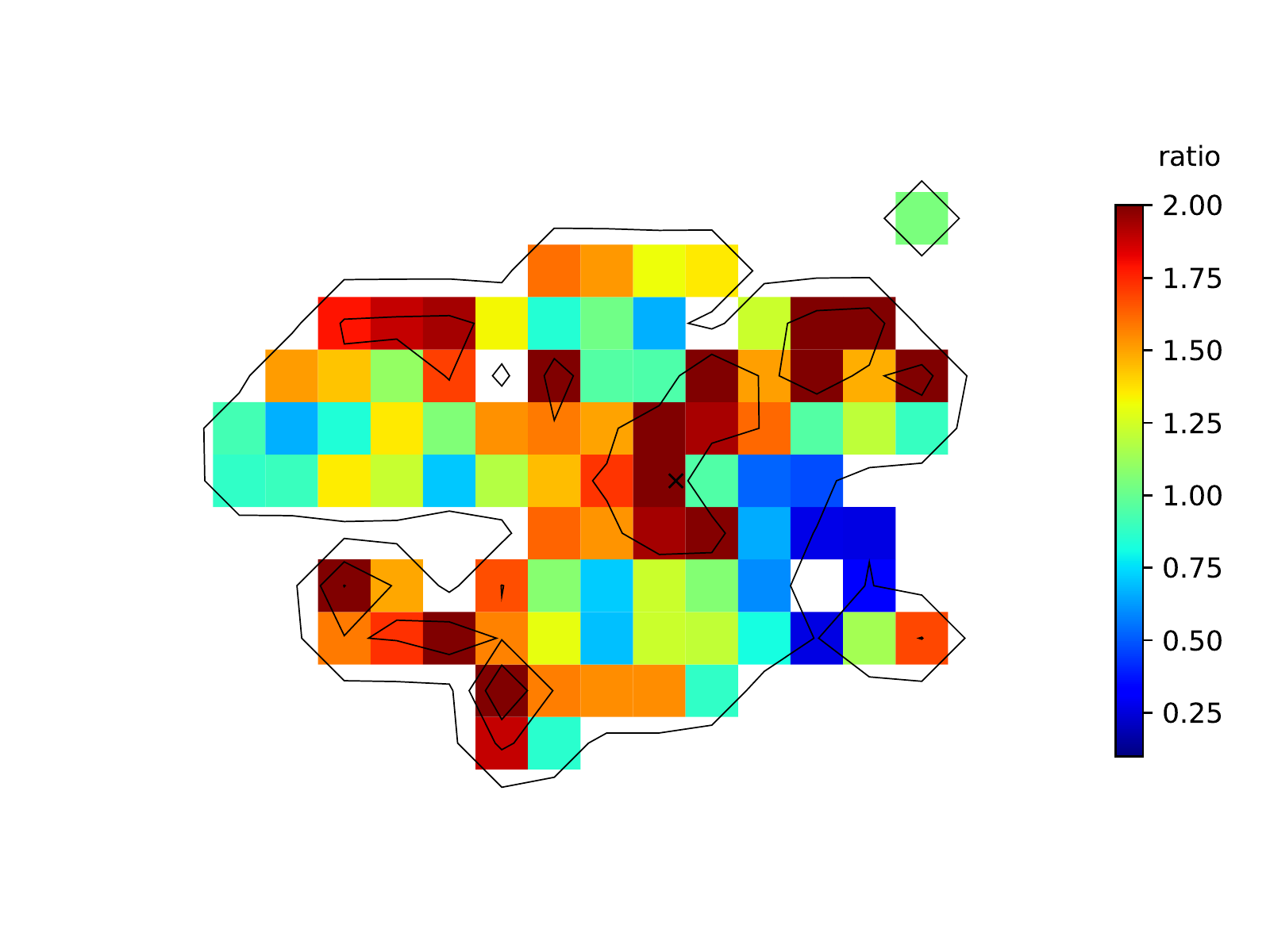}
\includegraphics[width=0.8\columnwidth]{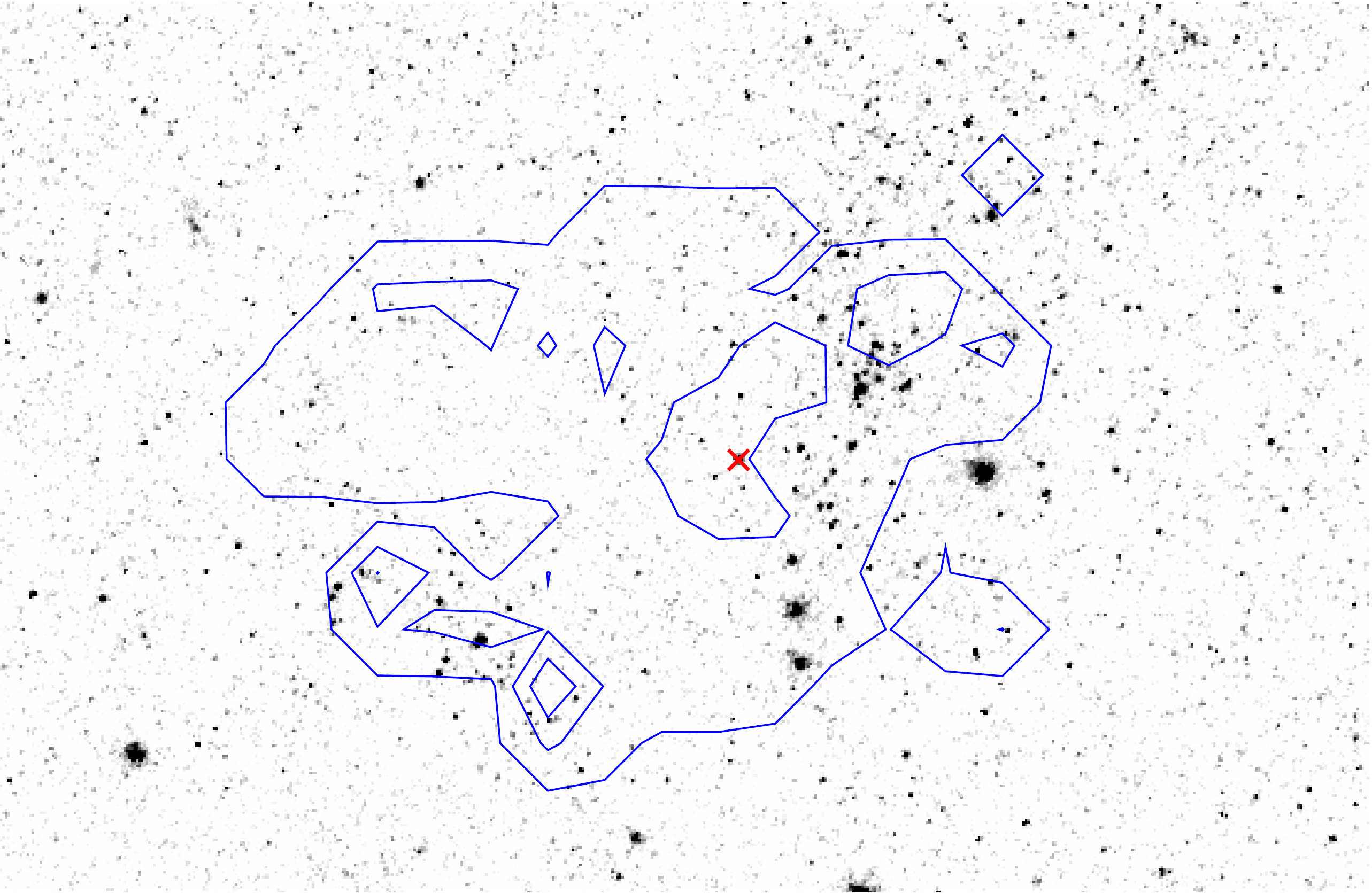}
\includegraphics[width=0.8\columnwidth]{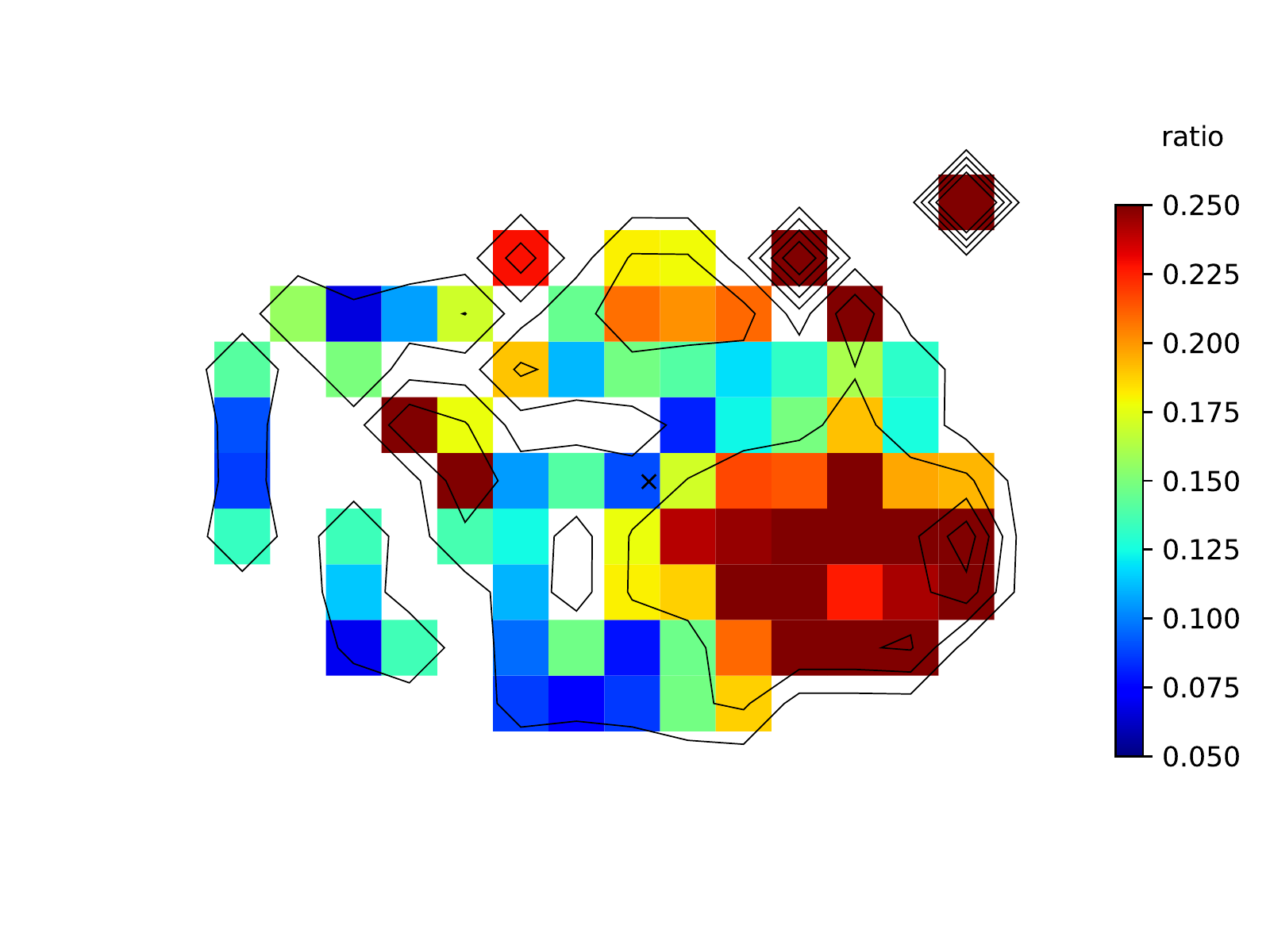}
\caption{\OIIItohb\ ratio map ({\bf top}) and its contours on top of the HST F435W image ({\bf middle}). The ratio map has a binning of $9 \times 9$ and only shows pixels with a signal-to-noise ratio greater than 2. The cross marks the ULX position. In the {\bf bottom} panel, the \OIwave$/$\halpha\ ratio map is displayed for comparison; it shows an anti-correlation with the \OIIItohb\ map.
\label{fig:ratio}}
\end{figure}

In Figure~\ref{fig:ratio}, we show the \OIIItohb\ ratio map on top of the HST F435W image.  Regions with enhanced \OIIItohb\ need more contribution from photoionization, including the ULX region and the southeastern and northwestern parts of the nebula. For comparison, the \OIwave$/$\halpha\ map is also shown in Figure~\ref{fig:ratio}. The distribution of \OIwave$/$\halpha\ is anti-correlated with that of \OIIItohb, as they require oxygens at different ionization states. Particularly, the southwestern region shows high \OIwave$/$\halpha\ and low \OIIItohb. We try to model the emission line ratios in this region with Model B, and find that less X-ray luminosity ($7 \times 10^{39}$~\ergs) and less number (20) of O stars are needed, compared with that needed for the whole nebula (see Table~\ref{tab:ratio}), indicative of a lower pre-ionization state in this region.

\subsection{Total power}

Following \citet{Pakull2002}, here we estimate the mechanical power of the wind/jet that inflates the bubble.  The \hbeta\ surface brightness ($S_{\rm H\beta}$) is a function of the shock velocity ($v_{\rm s}$) and the pre-shock ISM number density \citep[$n_{\rm ISM}$;][]{Dopita1996}, 
\begin{equation}
\frac{S_{\rm H\beta}}{{\rm erg \, s}^{-1} \, {\rm cm}^{-2}} = 7.44 \times 10^{-6} \left( \frac{{v_{\rm s}}}{10^2\, {\rm km\, s}^{-1}} \right)^{2.41} \left( \frac{n_{\rm ISM}}{{\rm cm}^3} \right) \, .
\end{equation}
Observationally, the surface brightness can be calculated via the line luminosity ($L_{\rm H\beta}$) or observed line flux ($f_{\rm H\beta}$),
\begin{equation}
S_{\rm H\beta} = \frac{L_{\rm H\beta}}{4 \pi R_{\rm B}^2} = \frac{f_{\rm H\beta}}{\theta_{\rm B}^2} \; ,
\end{equation}
where $R_{\rm B}$ and $\theta_{\rm B}$ are the physical radius and angular radius, respectively, of the nebula. Here we take the size of the semi-major axis of the nebula into calculation.  We take a shock velocity $v_{\rm s} = 80$~\kms\ and assume Galactic extinction only.  Thus, for the whole nebula, we obtain an average $n_{\rm ISM} = (0.45 \pm 0.02)$~cm$^{-3}$.

Assuming a pressure driven nebula \citep{Weaver1977},  the wind/jet power ($P$) and the age of the nebula ($t$) are related with other properties of the nebula as follows,
\begin{equation}
R_{\rm B} \approx 0.76 \, t^\frac{3}{5} \left( \frac{P}{\rho_0} \right)^\frac{1}{5} \; ,
\end{equation}
\begin{equation}  
v_{\rm s} \approx 0.38 \left( \frac{P}{R_{\rm B}^2 \rho_0} \right)^\frac{1}{3} \; , \; \rm{and}
\label{eq:vs}
\end{equation}
\begin{equation}
t = \frac{3R_{\rm B}}{5v_{\rm s}} \; ,
\end{equation}
where $\rho_0 = \mu m_{\rm p} n_{\rm ISM}$ is the pre-shock density of the ISM, $m_{\rm p}$ is the proton mass, and $\mu \approx 1.2$ is the average atomic weight assuming neutral gas with $Z = 0.5 Z_\sun$.  Plugging in $R_{\rm B}$ and $n_{\rm ISM}$ and considering the measurement uncertainties, we get
\begin{equation}
t \approx 2.1 \; {\rm Myr,}
\end{equation}
and
\begin{equation}
P = (6.5 \pm 0.3) \times 10^{39} \; {\rm erg\, s}^{-1} \; .
\end{equation}
All these results are generally consistent with those reported previously \citep[e.g.,][]{Pakull2002}.

\section{Discussion}
\label{sec:dis}

With the MUSE data, we obtain results consistent with previous studies for the whole nebula. The flux, velocity, velocity dispersion maps help reveal interesting structures inside the bubble.  The surface brightness is scaled with shock velocity and ISM density. Therefore, the low surface brightness in region 3 can be explained as a result of a low ISM density.  

In region 2 (ULX region), where the gas bulk motion is radial, we find that a simple two-component (approaching and receding) model cannot fit the line profile, and a third low-velocity narrow component is needed. Such a component could be a hint of photoionization.  We find spatial correlation between the \OIII\ emission and clusters of stars, as well as enhanced \OIII\ emission in region 2 around the ULX, suggesting that some of the \OIII\ emission may be due to photoionization. Simulations with \emph{MAPPINGS V} also suggest that pure shocks are unable to account for the observed line ratios; shocks with $\le$80~\kms\ cannot produce \OIIItohb\ as high as that observed, while shocks with higher velocities that are able to reproduce the observed \OIIItohb\ cannot explain the observed high value of \OI$/$\halpha\ due to over-ionization.  Adding photoionization to shocks of 80~\kms\ can reproduce the observations.  In that case,  the X-ray luminosity is derived to be close to $10^{40}$~\ergs, which is comparable to but slightly higher than the typical isotropic luminosity (several times $10^{39}$~\ergs) of NGC 1313 X-2 inferred from X-ray observations \citep{Qiu2021}. We note that both luminosities are derived assuming an illuminating distance of 300~pc. If the emission line clouds are located at a smaller distance to the source, a lower luminosity is required.  For X-rays, a reduced distance by a factor less than 2 can match the luminosity seen from X-ray observations. For O stars, as the brightest \OIII\ region is spatially coincident with star clusters,  they could be close in distance and only a few of them can account for the needed luminosity (e.g., one or two O stars at a distance of 50~pc). Therefore, the presence of these ionizing sources can account for the power needed for photoionization. However, the power-law component extrapolated from the X-ray band may cut off at energies above EUV, while the blackbody component observed in the soft X-ray band cannot provide sufficient ionization.  Alternatively, a flat ionizing spectrum with a higher X-ray luminosity ($5 \times 10^{40}$~\ergs) plus more O stars ($\sim$250), both at 300~pc, can provide a marginal fit.  If this is the case, it may suggest that the X-ray ionizing source is a multicolor accretion flow and the ionizing distance is smaller than that assumed, e.g.,  $1.4 \times 10^{39}$~\ergs\ at 50~pc.  We note that, although the total luminosity seems to fit the observation, one should keep in mind that the origin of the X-ray source for ionization is still uncertain. 

Additional photoionization may affect the the estimation of the wind/jet power that is based on a shock model \citep{Weaver1977}. However, the current data do not allow us to accurately quantify the luminosity and spectrum of the photoionization source. Also, the extra photoionization changes the pre-ionization state and has a nonlinear contribution to the production of emission lines. Therefore, we may treat the inferred power as an upper limit of the wind/jet power for the ULX.

\begin{acknowledgments}
We thank the anonymous referee for useful comments. HF acknowledges funding support from the National Key R\&D Project under grant 2018YFA0404502, the National Natural Science Foundation of China under grants Nos.\ 12025301 \& 11821303, and the Tsinghua University Initiative Scientific Research Program.
\end{acknowledgments}


\end{document}